\documentclass[slac_one]{revtex4}
\usepackage{graphicx}
\usepackage{fancyhdr}
\begin{document}

\newcommand{\be}{\begin{equation}}
\newcommand{\ee}{\end{equation}}
\newcommand{\bfm}[1]{\mbox{\boldmath$#1$}}
\newcommand{\bff}[1]{\mbox{\scriptsize\boldmath${#1}$}}
\newcommand{\al}{\alpha}
\newcommand{\bt}{\beta}
\newcommand{\lm}{\lambda}
\newcommand{\bea}{\begin{eqnarray}}
\newcommand{\eea}{\end{eqnarray}}
\newcommand{\gm}{\gamma}
\newcommand{\Gm}{\Gamma}
\newcommand{\dl}{\delta}
\newcommand{\Dl}{\Delta}
\newcommand{\ep}{\varepsilon}
\newcommand{\vep}{\varepsilon}
\newcommand{\kp}{\kappa}
\newcommand{\Lm}{\Lambda}
\newcommand{\om}{\omega}
\newcommand{\pa}{\partial}
\newcommand{\nn}{\nonumber}
\newcommand{\dd}{\mbox{d}}
\newcommand{\grtsim}{\mbox{\raisebox{-3pt}{$\stackrel{>}{\sim}$}}}
\newcommand{\lessim}{\mbox{\raisebox{-3pt}{$\stackrel{<}{\sim}$}}}
\newcommand{\uk}{\underline{k}}
\newcommand{\gsim}{\;\rlap{\lower 3.5 pt \hbox{$\mathchar \sim$}} \raise 1pt \hbox {$>$}\;}
\newcommand{\lsim}{\;\rlap{\lower 3.5 pt \hbox{$\mathchar \sim$}} \raise 1pt \hbox {$<$}\;}
\newcommand{\Li}{\mbox{Li}}
\newcommand{\bc}{\begin{center}}
\newcommand{\ec}{\end{center}}

\def\lapprox{\lower .7ex\hbox{$\;\stackrel{\textstyle <}{\sim}\;$}}
\def\gapprox{\lower .7ex\hbox{$\;\stackrel{\textstyle >}{\sim}\;$}}

\title{Fermionic Corrections to the Heavy-Quark Pair Production in the Quark-Antiquark Channel}

\author{R.~Bonciani, A.~Ferroglia, T.~Gehrmann, C.~Studerus}
\affiliation{University of Z\"urich, Z\"urich, CH 8057, Switzerland }
\author{D.~Ma\^itre}
\affiliation{SLAC, Stanford, CA 94025, USA}

\begin{abstract}
We describe the analytic calculation of the fermionic two-loop QCD
corrections to the heavy-quark pair production process in the
quark-antiquark channel.
\end{abstract}

\maketitle

\thispagestyle{fancy}

\section{INTRODUCTION}

The top quark is the heaviest fermion of the Standard Model. Since its discovery at
the Fermilab Tevatron~\cite{toptev}, its mass has  been measured to within a few
percent, while its production cross  section and couplings are currently known with
larger uncertainty.  With the large number of top quarks expected to be produced at
the  LHC, the study of its properties will become precision physics. To  interpret
these upcoming precision data, equally precise theoretical  predictions are
mandatory. These demand foremost the calculation  of higher order corrections in
perturbative QCD.

At present, the top quark pair production cross section is known to next-to-leading
order (NLO) in the QCD coupling constant~\cite{Nason:1987xz}. For this process, the
resummation of next-to-leading logarithmically enhanced  corrections (NLL) improves
upon the fixed-order NLO prediction~\cite{Kidonakis:1997gm}. Electroweak one-loop
corrections to $t\bar t$ production are equally available \cite{Kuhn:2005it}.
For the top quark pair production cross section, which is expected to be measured to
within a few percent accuracy, the currently available theoretical prediction is not sufficiently
precise. Recent studies~\cite{Moch:2008qy} indicate a scale uncertainty on these
predictions of 7\%, and a parton distribution uncertainty of 6\%. While the latter
may be improved upon by more precise determinations of the parton distribution
functions at HERA and LHC, the former requires the calculation of perturbative
corrections at next-to-next-to-leading order (NNLO) in QCD.

The calculation of the full NNLO corrections to the top quark pair production cross
section requires three types of ingredients: two-loop matrix elements for $q\bar q
\to t\bar t$ and $gg\to t\bar t$, one-loop matrix elements for hadronic production
of $t\bar t+$(1 parton)  and tree-level matrix elements for hadronic production of
$t\bar t+$(2 partons). The latter two ingredients were computed previously in the
context of the NLO corrections to $t\bar t$+jet production~\cite{Dittmaier:2007wz}.
They contribute to the  $t\bar t$ production cross section through configurations
where up to two final state partons can be unresolved (collinear or soft), and their
implementation thus may require further developments of subtraction techniques at
NNLO.

Both two-loop matrix elements were computed analytically in the small-mass expansion
limit $s, |t|, |u| \gg m^2$ in~\cite{Czakon:2007ej}, starting from the previously
known massless two-loop matrix elements for $q\bar q\to q'\bar
q'$~\cite{Anastasiou:2000kg} and $gg\to q\bar q$~\cite{Anastasiou:2001sv}. An exact
numerical representation of the two-loop matrix element  $q\bar q \to t\bar t$ has
been obtained very recently~\cite{Czakon:2008zk}. In \cite{US} we computed all
two-loop contributions to $q\bar q \to t\bar t$ arising from closed fermion loops in
a compact analytic form, providing also a first independent validation of the
results of~\cite{Czakon:2007ej,Czakon:2008zk}. Our results allow for a fast
numerical evaluation and permit the analytical study of the cross section near
threshold. In the rest of this proceeding we
briefly discuss the structure of the two-loop fermionic corrections and
the calculational techniques employed to evaluate them.
%in \cite{US}.

\section{STRUCTURE}

The scattering process we consider is $q(p_1) + \overline{q}(p_2)
\to  t(p_3) + \overline{t}(p_4)$
in Euclidean kinematics, where $p_i^2 = 0$ for $i=1,2$  and $p_j^2
= -m^2$ for  $i=3,4$. The Mandelstam variables are defined as
follows:
$s = -\left(p_1 + p_2 \right)^2$, $t = -\left(p_1 - p_3
\right)^2$, $u = -\left(p_1 - p_4 \right)^2$.  Conservation of
momentum implies that $s +t +u = 2 m^2$.
The squared matrix element (averaged over the spin and color of
the incoming quarks and summed over the spin of the outgoing
ones), calculated in $d = 4 -2 \varepsilon$ dimensions, can be
expanded in powers of the strong coupling constant $\alpha_S$ as
follows:
\be \label{M2} |\mathcal{M}|^2(s,t,m,\varepsilon) = \frac{4 \pi^2
\alpha_S^2}{N_c^2} \left[{\mathcal A}_0 + \left(\frac{\alpha_s}{
\pi} \right) {\mathcal A}_1 + \left(\frac{\alpha_s}{ \pi}
\right)^2 {\mathcal A}_2 + {\mathcal O}\left(
\alpha_s^3\right)\right] \, . \ee
The tree-level amplitude involves a single diagram
 and its contribution to Eq.~(\ref{M2}) is
given by
\be
{\mathcal A}_0=  4  N_c \, C_F \left[
\frac{(t-m^2)^2+(u-m^2)^2}{s^2} + \frac{2 m^2}{s} -
\varepsilon \right] \, ,
\ee
where  $N_c$ is the number of colors and  $C_F = (N_c^2-1)/2N_c$.

The NLO term ${\mathcal A}_1$  in Eq.~(\ref{M2}) arises from the interference of
one-loop diagrams with the tree-level amplitude \cite{Nason:1987xz}.
The NNLO term ${\mathcal A}_2$ consists of two parts, the interference of two-loop
diagrams with the Born amplitude and the interference of one-loop diagrams among
themselves: $ {\mathcal A}_2 = {\mathcal A}_2^{(2\times 0)} + {\mathcal
A}_2^{(1\times 1)}$. The latter term ${\mathcal A}_2^{(1\times 1)}$ was studied
extensively  in \cite{Korner:2005rg}. ${\mathcal A}_2^{(2\times 0)}$
%originating
%from the two-loop  diagrams,
can be decomposed according to color and flavor
structures as follows:
\bea
{\mathcal A}_2^{(2\times 0)}  &=&  N_c C_F \left[ N_c^2 A
+B +\frac{C}{N_c^2}  + N_l \left( N_c D_l + \frac{E_l}{N_c}
\right)
+ N_h \left( N_c D_l + \frac{E_l}{N_c} \right)
+ N_l^2 F_l + N_l N_h F_{lh} + N_h^2 F_h
\right] \, ,
\label{colstruc}
\eea
where $N_l$ and $N_h$ are the number of light- and heavy-quark flavors,
respectively. The coefficients $A,B,\ldots,F_h$ in Eq.~(\ref{colstruc}) are
functions of $s$, $t$, $m$, and
$\varepsilon$. These quantities were calculated in  \cite{Czakon:2007ej} in the
approximation $s,|t|,|u| \gg m^2$. For a fully differential description of top quark
pair production at NNLO, the complete mass dependence of  ${\mathcal A}_2^{(2\times
0)}$ is required. An exact numerical expression for it has been obtained in
\cite{Czakon:2008zk}. In \cite{US}, we derived exact analytic expressions for all
the terms in Eq.~(\ref{colstruc}) arising from two-loop diagrams involving at least
a fermion loop (i.e. the coefficients $D_i,E_i,F_j$ with $i=l,h$ and $j=l,h,lh$),
providing also an independent confirmation of the  results of
\cite{Czakon:2007ej,Czakon:2008zk}.

\section{CALCULATION\label{sec:calc}}

The two-loop Feynman diagrams for $q\bar q\to t\bar t$ were generated with
QGRAF~\cite{qgraf}. The interference with the tree-level amplitude, as well as the
color and Dirac algebra, were simplified by using a FORM~\cite{FORM} code. Out of
the $\sim$ 200 two-loop diagrams contributing to the amplitude, about 60 are
proportional to $N_l$ and/or $N_h$. There is only one two-loop box topology
contributing to the $N_l$ part of the squared amplitude, and a single other two-loop
box topology proportional to $N_h$. These two box topologies are very similar to the
ones encountered in the evaluation of the two-loop QED corrections to Bhabha
scattering \cite{electronloop,hfbha}, and can be evaluated with the same techniques.

All two-loop integrals appearing in these amplitudes are reduced to a set of master
integrals (MIs) using two independent implementations of the Laporta algorithm
\cite{Laportaalgorithm}.
Only part of these MIs were available in the literature~\cite{Fleischer:1999hp}
from previous two-loop calculations of the heavy quark form factors
\cite{Bernreuther:2004ih} and amplitudes for Bhabha scattering
\cite{electronloop,hfbha,bhabha}. The remaining MIs
%, including four two-loop
%four legs integrals,
were evaluated in \cite{US} by employing the differential equation method
\cite{DiffEq}.

All the MIs were calculated in the non-physical region $s<0$.
The transcendental functions appearing in the MIs are one- and two-dimensional
harmonic polylogarithms (HPLs) \cite{HPLs} of maximum weight four and three,
respectively. Both sets of functions can be rewritten in  terms of conventional
Nielsen's polylogarithms.

Following the procedure outlined in the present section, it was possible to obtain
the expression of the bare squared matrix elements involving diagrams proportional
to $N_l$ and/or $N_h$. The UV divergencies were renormalized in a mixed scheme
described in detail in \cite{US}.
In order to cross check our analytical results, we expanded them in the $s,|t|, |u|
\gg m^2$ limit. The first term in the expansion agrees with the results published in
\cite{Czakon:2007ej}; the second order term agrees with the results found in the
{\tt Mathematica} files  included in the arXiv version of \cite{Czakon:2008zk}. We
also find complete agreement with the numerical result of Table~3 in
\cite{Czakon:2008zk}, corresponding to a phase  space point in which the $s,|t|, |u|
\gg m^2$ approximation cannot be applied.

\begin{acknowledgments}

Work supported by the Swiss National Science Foundation (SNF)
under contract 200020-117602. 
\end{acknowledgments}

\end{document}